\documentclass[11pt]{article}
\pdfoutput=1
\usepackage{jcapmod}
\usepackage{booktabs}
\usepackage[english]{babel}
\usepackage{amsmath,amssymb,amsbsy,amstext, amsthm, simplewick}
\usepackage{graphicx}
\usepackage{amsfonts}
\usepackage{amssymb}
\usepackage{upgreek}
\usepackage{exscale,relsize}
\usepackage[makeroom]{cancel}
\usepackage{soul}
\usepackage{slashed}

\RequirePackage{color}

\usepackage{colortbl}
\definecolor{rp}{cmyk}{0.2, 1, 0.6, 0}
\definecolor{green2}{cmyk}{0, 1, 0.5, 0}
\definecolor{lightgreen}{cmyk}{0.2, 0, 0.2, 0.2}
\definecolor{lightgray}{cmyk}{0.1,0.2,0,0.1}
\definecolor{lightgray2}{cmyk}{0.4,0.4,0,0.8}
\definecolor{black}{cmyk}{1.0,1.0,1.0,1.0}

\allowdisplaybreaks[1]

\usepackage{colortbl}
\definecolor{lightgreen}{cmyk}{0.2, 0, 0.2, 0.2}
\definecolor{lightgray}{cmyk}{0.1,0.2,0,0.1}
\definecolor{lightgray2}{cmyk}{0.1,0.1,0,0.1}

\makeatletter
\newlength{\apb@width}
\newcommand{\autoparbox}[2][c]{\settowidth{\apb@width}{#2}\parbox[#1]{\apb@width}{#2}}

\makeatother

\numberwithin{equation}{section}

\def\beq{\begin{equation}}
\def\eeq{\end{equation}}

\def\bea{\begin{eqnarray}}
\def\eea{\end{eqnarray}}

\def\beq{\begin{equation}}
\def\eeq{\end{equation}}
\def\bea{\begin{eqnarray}}
\def\eea{\end{eqnarray}}

\def\Neff{N_{\rm eff}}
\def\pann{p_{\rm ann}}

\def\0{{\boldsymbol 0}}

\def\pann{{p_{\rm ann}}}

\DeclareRobustCommand{\SkipTocEntry}[4]{}

\begin{document}

\begin{titlepage}

\setcounter{page}{1} \baselineskip=15.5pt \thispagestyle{empty}

\bigskip\

\vspace{1cm}
\begin{center}

{\fontsize{20}{28}\selectfont  \sffamily \bfseries Aspects of Dark Matter  \\[10pt] Annihilation in Cosmology}

\end{center}

\vspace{0.2cm}

\begin{center}
{\fontsize{14}{30}\selectfont   Daniel Green$^1$, P. Daniel Meerburg$^{2,3,4,5}$ and Joel Meyers$^6$ }
\end{center}

\begin{center}
\textsl{$^1$ Department of Physics, University of California, San Diego, La Jolla, CA 92093, USA} \\
\textsl{$^2$ Kavli Institute for Cosmology, Madingley Road, Cambridge, UK, CB3 0HA} \\
\textsl{$^3$ DAMTP, Centre for Mathematical Sciences, Wilberforce Road, Cambridge, UK, CB3 0WA} \\
\textsl{$^4$ Kapteyn Astronomical Institute, University of Groningen, P.O. Box 800, 9700 AV Groningen, The Netherlands} \\
\textsl{$^5$ Van Swinderen Institute for Particle Physics and Gravity,\\ University of Groningen,
Nijenborgh 4, 9747 AG Groningen, The Netherlands} \\
\textsl{$^6$ Canadian Institute for Theoretical Astrophysics, Toronto, ON M5S 3H8, Canada} \\
\
\end{center}

\vspace{1.2cm}
\hrule \vspace{0.3cm}
\noindent {\sffamily \bfseries Abstract} \\[0.1cm]
Cosmic microwave background (CMB) constraints on dark matter annihilation are a uniquely powerful tool in the quest to understand the nature of dark matter.  Annihilation of dark matter to Standard Model particles between recombination and reionization heats baryons, ionizes neutral hydrogen, and alters the CMB visibility function.  Surprisingly, CMB bounds on dark matter annihilation are not expected to improve significantly with the dramatic improvements in sensitivity expected in future cosmological surveys.  In this paper, we will present a simple physical description of the origin of the CMB constraints and explain why they are nearly saturated by current observations.  The essential feature is that dark matter annihilation primarily affects the ionization fraction which can only increase substantially at times when the universe was neutral.  The resulting change to the CMB occurs on large angular scales and leads to a phenomenology similar to that of the optical depth to reionization.  We will demonstrate this impact on the CMB both analytically and numerically.  Finally, we will discuss the additional impact that changing the ionization fraction has on large scale structure.

\vskip 10pt
\hrule

\vspace{0.6cm}
 \end{titlepage}

 \tableofcontents

\newpage

\section{Introduction}

Dark matter annihilation is a universal prediction of the weakly interacting massive particles (WIMP) paradigm.  The dark matter annihilation cross-section determines the relic abundance and thus offers a compelling target for astrophysical and cosmological observations.  Cosmic microwave background (CMB) constraints on dark matter annihilation have been enormously valuable given the vast space of possible models, since these constraints are mostly insensitive to the details of the dark matter model. The current constraint derived from data collected by the Planck satellite requires $\pann <3.4\times 10^{-28} \, \mathrm{cm}^3/\mathrm{s}/\mathrm{GeV}$ (95\% CL)~\cite{Ade:2015xua}, where $\pann = f_\mathrm{eff} \frac{\langle \sigma v \rangle}{m_\chi}$ and $f_\mathrm{eff}$ is a parameter of order unity describing how the amount of baryon heating depends on annihilation products (see e.g.~\cite{Slatyer:2009yq}).  This constraint excludes the WIMP thermal cross section for dark matter masses $m_{\chi}<{\cal O}(10)$ GeV and puts pressure on many models with detectable astrophysical signatures (see e.g.~\cite{Slatyer:2017sev} for a review). The CMB constraint on $\pann$ is particularly valuable because it is both a clean measurement and is gleaned from the same system that we use to measure the dark matter abundance itself.  

\begin{figure}[t]
\includegraphics[width=\columnwidth]{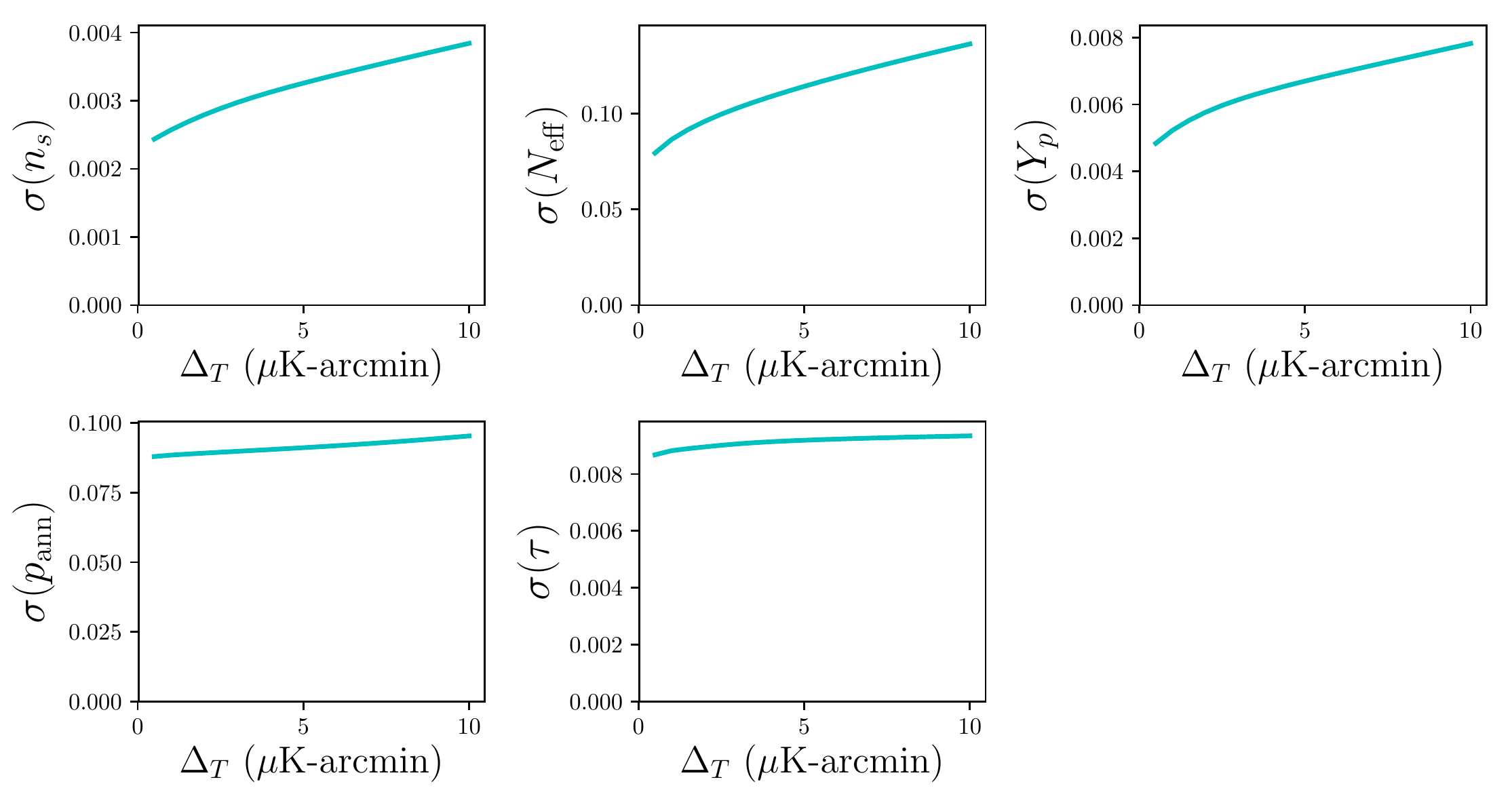}
\caption{ Forecasted 1-$\sigma$ errors on selected parameters for CMB experiments over a range of noise levels, assuming a 1-arcmin beam, $\ell_{\mathrm{min}} = 30$, $\ell_\mathrm{max}^{TT} = 3000$, and $\ell_\mathrm{max} = 5000$, with $f_\mathrm{sky} = 0.4$, including also a Planck-like experiment on large scales and for additional sky coverage.  The constraints on $\pann$ are shown in units of $10^{-27} \,  \mathrm{cm}^3/\mathrm{s}/\mathrm{GeV}$.  There is only marginal reduction in the forecasted errors on $p_\mathrm{ann}$ (and $\tau$) across this range of noise levels, while the other parameters show marked improvement. }
\label{fig:forecasts}
\end{figure}

Despite the importance of the CMB constraint on dark matter annihilation, we lack a complete understanding of the origin of the constraint and the fundamental limitations of the measurement. The strength of the CMB constraint is often explained by noting that the annihilation rate scales as the square of the dark matter density. A priori, one might then conclude that the CMB constraint is dominated by dark matter annihilation near the time of recombination, where the dark matter density is largest.  The problem with this interpretation is that it fails to explain the evolution of the constraint on $\pann$ over the past decade and the behavior of forecasted future constraints. The coming generation of CMB experiments is poised to make an order of magnitude improvement on parameters known to affect the physics of recombination, such as $\Neff$ and $Y_p$ \cite{Abazajian:2016yjj}.  While dark matter annihilation also visibly affects the power spectra on small angular scales, the forecasts (Figure~\ref{fig:forecasts}) show effectively no improvement for future observations.  Such behavior could suggest a degeneracy with another cosmological parameter, such as $n_s$.  Unfortunately,\footnote{The presence of such a degeneracy would suggest a strategy for improving constraints on $\pann$ by measuring the degenerate cosmological parameters through other observables (like galaxy clustering).} forecasts do not exhibit any clear degeneracy with the parameters of the $\Lambda$CDM model or extensions thereof. These results suggest instead that the information in the CMB has been largely saturated by current observations.

In this paper, we will offer a simple physical explanation for these curious trends seen in the forecasts.  We will review how modifications to the physics of recombination change the appearance of the CMB and demonstrate that $\pann$ forecasts do not exhibit the same characteristics. Instead, we will see that $\pann$ behaves (suggestively) most like the optical depth to reionization, $\tau$.  We will then show, both analytically and numerically, that the constraints arise primarily from annihilation at redshifts {\it between} recombination and reionization. The dominant effect of dark matter annihilation is to increase the ionization fraction, $x_e$, which can only change significantly when the universe is nearly neutral. Furthermore, ionization of hydrogen due to dark matter annihilation is balanced by the capture of free electrons, which scales as the square of the electron density and thus offsets the larger annihilation rate at earlier times.  This balance leads to an increased and nearly constant ionization fraction between recombination and reionization, thereby altering the CMB visibility function during this epoch.  Due to the low-redshift character of the effects of dark matter annihilation, the observational imprints occur on large angular scales, thus explaining the similarity of constraints on $\pann$ and $\tau$ in forecasts.  

The behavior seen in forecasts was noted in previous studies, such as~\cite{Padmanabhan:2005es, Galli:2009zc, Madhavacheril:2013cna,Slatyer:2015jla}. In particular, low-$\ell$ E-mode measurements are known to produce the strongest bounds on dark matter annihilation. Principal component analysis also revealed that CMB constraints are particularly sensitive to annihilation at $z\approx 600$ \cite{Finkbeiner:2011dx}. We will see that this is not a unique feature of the \emph{polarization} spectra, as $\pann$ also impacts \emph{temperature} spectra in a manner similar to the optical depth, $\tau$. Furthermore, we will show this phenomenology can be captured with a fairly simple physical model.

Given the impact on the low(er) redshift universe, it is natural to investigate how dark matter annihilation alters the evolution of large scale structure. We will show that there is an additional effect unique for large scale structure due to increased photon drag. Unfortunately this effect, which only impacts baryon fluctuations, is suppressed in the baryon fluctuations at low redshift due to the gravitational influence of the dark matter.  The effect of increased photon drag could in principle provide some additional constraining power on dark matter annihilation from studies of the early baryonic universe, which could be observed indirectly through 21cm absorption in the dark ages.

This paper is organized as follows: In Section~\ref{sec:highell}, we will discuss the phenomenology and forecasts associated with the physics of recombination.  We will see  $\pann$ does not neatly fit into this description and instead is most similar to $\tau$ in forecasts.  In Section~\ref{sec:pheno}, we will provide an analytic description of how $\pann$ alters the CMB and match these expectation with numerical spectra.  In Section~\ref{sec:lss}, we will study the impact on baryons at low redshifts.  In Section~\ref{sec:sum} we conclude.  Appendix~\ref{app:degen} provides additional details regarding our forecasts.

\section{CMB Physics at Small Angular Scales}\label{sec:highell}

In the early universe, dark matter annihilation heats the baryon-photon fluid and ionizes hydrogen atoms.  The natural question of interest is how these changes manifest themselves in observable quantities like the temperature and polarization fluctuations in the CMB.  
In this section, we will give a phenomenological description of the physics at small angular scales focusing on the physics of recombination in particular.  We will show that our phenomenological description accurately predicts the behavior of $Y_p$ and $\Neff$ but not $\pann$.  We will explore the physical reasons behind this in the next section. 

\subsection{Physics of the Damping Tail}

The CMB power on small angular scales is sensitive to the nature of the photon-baryon plasma around the era of recombination.  
With better angular resolution and lower noise than the Planck satellite, ground-based CMB experiments will make dramatic improvements measurement of the high-$\ell$ power spectra (both temperature and polarization)~\cite{Abazajian:2016yjj}.  The exponential damping on small angular scales is the dominant feature of these spectra that will drive cosmological constraints.  It is therefore useful to review the origin of this damping and how it is sensitive to cosmological parameters.  

The primary temperature or polarization fluctuations we observe can be compactly written (using the conventions of~\cite{Seljak:1996is}) as
\beq
\Delta_{T,P}(k,\mu,\eta_0) = \int^{\eta_0}_0 \mathop{d\eta} e^{i k\mu (\eta-\eta_0)} g(\eta) s_{T,P}(k,\eta) \, ,
\label{eq:DeltaTP}
\eeq
where $\eta$ is conformal time, $\mu=\mathbf{k}\cdot\mathbf{\hat{n}}/k$ for a given line of sight $\mathbf{\hat{n}}$,  and $g=\dot \tau e^{-\tau}$ is the visibility function.  The temperature and polarization source functions, $s_T$ and $s_P$,  are given by 
\bea
s_T &\approx& \Delta_{T,0} + \psi - \dot v_b - \frac{\Pi}{4} - \frac{4\ddot\Pi}{4 k^2} \, ,  \\
s_P &\approx& -\frac{3}{4 k^2} (k^2 \Pi + \ddot \Pi) \, , \\
\Pi &\equiv& \ \Delta_{T,2}+\Delta_{P,2}+\Delta_{P,0} \, .
\eea
The optical depth, $\tau$, is determined from the Thomson cross section, $\sigma_T$, through
\beq
\dot \tau = a n_e x_e \sigma_T,
\eeq
where $n_e$ is the number density of electrons and $x_e$ is the ionization fraction\footnote{We will define $n_e$ and $x_e$ ignoring electrons bound in helium such that $n_e = n_p +n_H$ and $x_e = n_p / (n_p + n_H)$.  This definition implies that $x_e$ can exceed unity at very early times before helium recombination and also at recent times after helium reionization, though neither epoch will be important for our discussion.}.

A common phenomenological description of recombination (see e.g.~\cite{Weinberg:2008zzc}) treats the visibility function as a Gaussian, such that
\beq
g(\eta) \approx \frac{1}{\sqrt{2 \pi \sigma_\star}} e^{-\frac{(\eta-\eta_\star)^2}{2 \sigma_\star^2}} \, ,
\eeq
where $\eta_\star$ is conformal time at recombination and $\sigma_\star \ll \eta_\star$ is the width of recombination. Using the tight-coupling approximation, the dominant contribution to the temperature source function is given by
\beq
(\Delta_{T,0} + \psi)(k,\eta) = \cos( k r_s(\eta)) e^{-k^2/k_D(\eta)^2} \, , 
\eeq
where $r_s(\eta)=\int^\eta d\eta' c_s(\eta')$ is the local sound horizon and 
\beq
\frac{1}{k_D^2} = \int^\eta \mathop{d\eta'} \frac{1}{\dot \tau} \frac{c_s^2}{2} \left[\frac{R_b^2}{(1+R_b)} + \frac{16}{15}\right]\ ,
\eeq
with $R_b = 3\rho_b/4\rho_r$. This source of exponential damping is due to diffusion of the photons and is related to the free-electron fraction via $\dot\tau$~\cite{Silk:1967aha}. 
We can integrate over $\eta$ in Eq.~\eqref{eq:DeltaTP} using the Gaussian approximation for the visibility function to find 
\beq\label{eq:damp}
\Delta_{T,P} \approx e^{i k\mu (\eta_\star-\eta_0)}\times \cos( k r_s(\eta_\star)) e^{-k^2(1/k_D(\eta_\star)^2 +1/k_L^2)} \, ,
\eeq
which contains an additional source of exponential damping due to fluctuations being averaged out over the finite width of recombination.  This Landau damping is given by
\beq
\frac{1}{k_L^2} = \frac{c_s^2 \sigma_\star^2}{2} = \frac{\sigma_\star^2}{6 (1+R_b)} \, .
\eeq
We see that the exponential suppression of the high-$k$ modes gets a contribution proportional to $x_e^{-1}$ from diffusion and one proportional to the width of recombination, $\sigma_\star^2$. Since we have evaluated these integrals around the time of recombination, we can use the Limber approximation to relate $\ell \approx k \eta_\star$.

\begin{figure}
\begin{centering}
\includegraphics[width=\columnwidth]{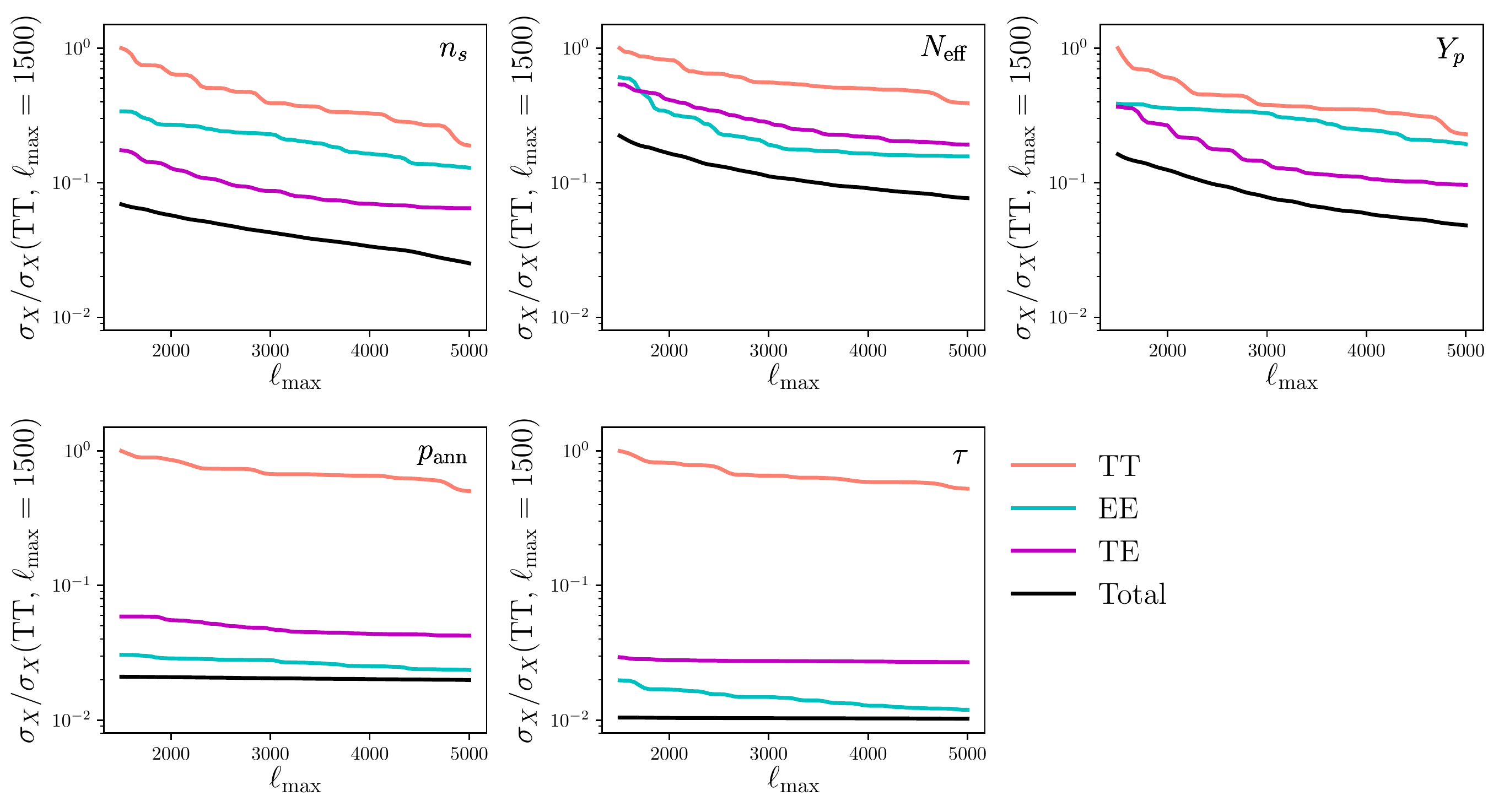}
\caption{ Improvement to 1-$\sigma$ errors on select parameters for a cosmic-variance-limited CMB experiment for various choices of $\ell_\mathrm{max}$, relative to a cosmic-variance-limited temperature-only experiment with $\ell_\mathrm{max} = 1500$.}
\label{fig:fisher_lmax}
\end{centering}
\end{figure}

\subsection{Impact of Cosmological Parameters}

From our phenomenological description of the high-$\ell$ behavior of the CMB, we can recover a reasonably clear understanding of how current and future measurements impact constraints on cosmological parameters which affect the epoch of recombination.

The helium fraction, $Y_p$, and the effective number of neutrino species, $\Neff$, are known to impact the high-$\ell$ CMB spectra. Accounting for degeneracies~\cite{Hou:2011ec}, the dominant impact of both parameters is to change $k_D$. Since helium has a higher binding energy, it contributes to the baryon density but recombines earlier than hydrogen and absorbs free electrons. Varying $Y_p$, while holding the physical baryon density $\omega_b$ fixed\footnote{Here we are assuming that there is no ionized helium around the time of hydrogen recombination, and we are neglecting the difference between the baryon number and atomic mass of helium. Shortly before hydrogen recombination, helium is entirely neutral, hydrogen is fully ionized, and $x_e \approx 1$ is unaffected by $Y_p$.  Under these assumptions, changing $Y_p$ at fixed $\omega_b$ changes the number free protons, $n_p$, and hence the number of free electrons, $n_e\sim n_p$.  At earlier times, the helium may be partially ionized which further modifies this formula because of the smaller the charge-to-mass ratio of helium.  } implies $n_e = n_b (1- Y_p)$. The contribution of light relics to the expansion rate changes the time of recombination $\eta_\star$, and when holding the angular size of the sound horizon $\theta_s$ fixed, one finds~\cite{Hou:2011ec}
\beq
\frac{1}{k_D^2} \propto (1+f_\nu(\Neff))^{0.56} (1-Y_p)^{-1},
\eeq
where 
\beq
f_\nu \equiv   \frac{\rho_\nu}{\rho_\gamma}= \frac{7}{8} \left(\frac{4}{11}\right)^{4/3}\Neff \ . 
\eeq
The measurement of $k_D$ from the high-$\ell$ CMB power spectra drives the constraint on either parameter individually, but when both parameters are included constraints degrade significantly~\cite{Bashinsky:2003tk,Baumann:2015rya}.

\begin{figure}
\begin{centering}
\includegraphics[width=1\columnwidth]{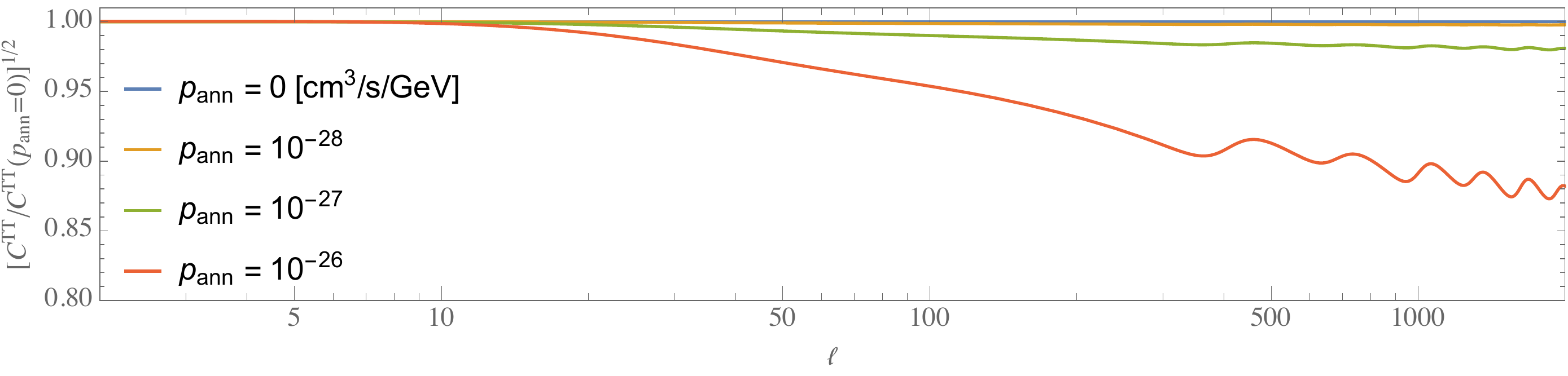}
\includegraphics[width=1\columnwidth]{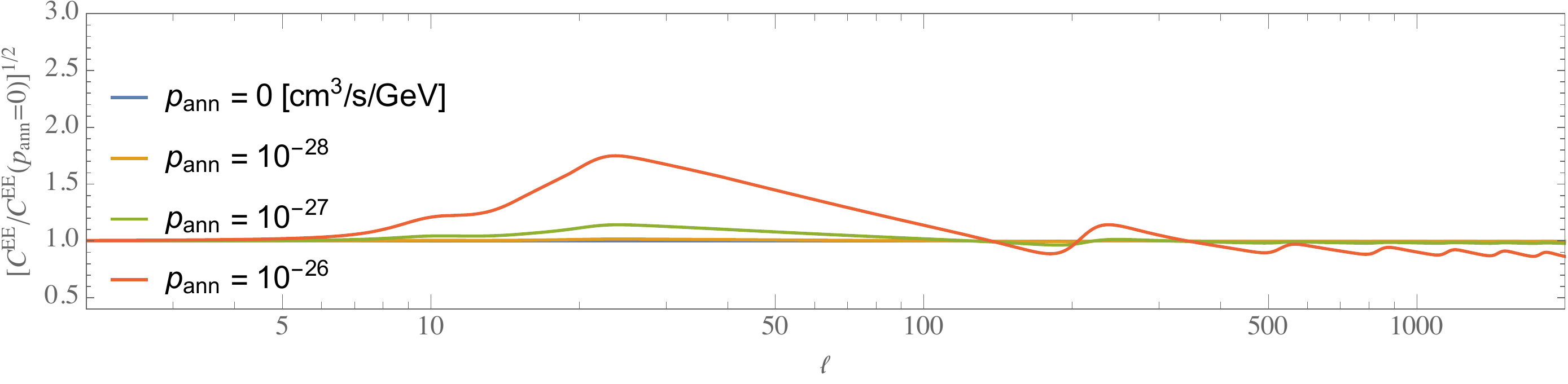}
\caption{The fractional change to the temperature (\emph{top}) and $E$-mode polarization (\emph{bottom}) power spectra due to dark matter annihilation.}
\label{fig:EffectOfDMAonTTEEXe}
\end{centering}
\end{figure}

Figures \ref{fig:forecasts} and~\ref{fig:fisher_lmax} show how constraints on $Y_p$ and $\Neff$ are expected to improve with better measurements of the small scale CMB power spectra.  These improvements are consistent with the idea that improved constraints come from a better measurement of the damping tail.  Beyond measuring the damping tail, polarization data can also help break degeneracies for a variety of well understood reasons~\cite{Bashinsky:2003tk,Baumann:2015rya}.  Nevertheless, we see significant improvements when more high-$\ell$ modes are included for every combination of spectra.

We now contrast this behavior with the forecasts for $\pann$.  At first glance, the change to the temperature power spectrum shown in Figure~\ref{fig:EffectOfDMAonTTEEXe} would suggest that the simple phenomenological picture presented in the previous section may also apply to dark matter annihilation. The heating from dark matter annihilation is expected to ionize neutral hydrogen and therefore it is natural to expect annihilation to increase the width of recombination, $\sigma_\star$, and hence $k_L$.  If true, Eq.~\eqref{eq:damp} shows we should expect behavior identical to $\Neff$ or $Y_p$.

\begin{figure}
\begin{centering}
\includegraphics[width=0.7\columnwidth]{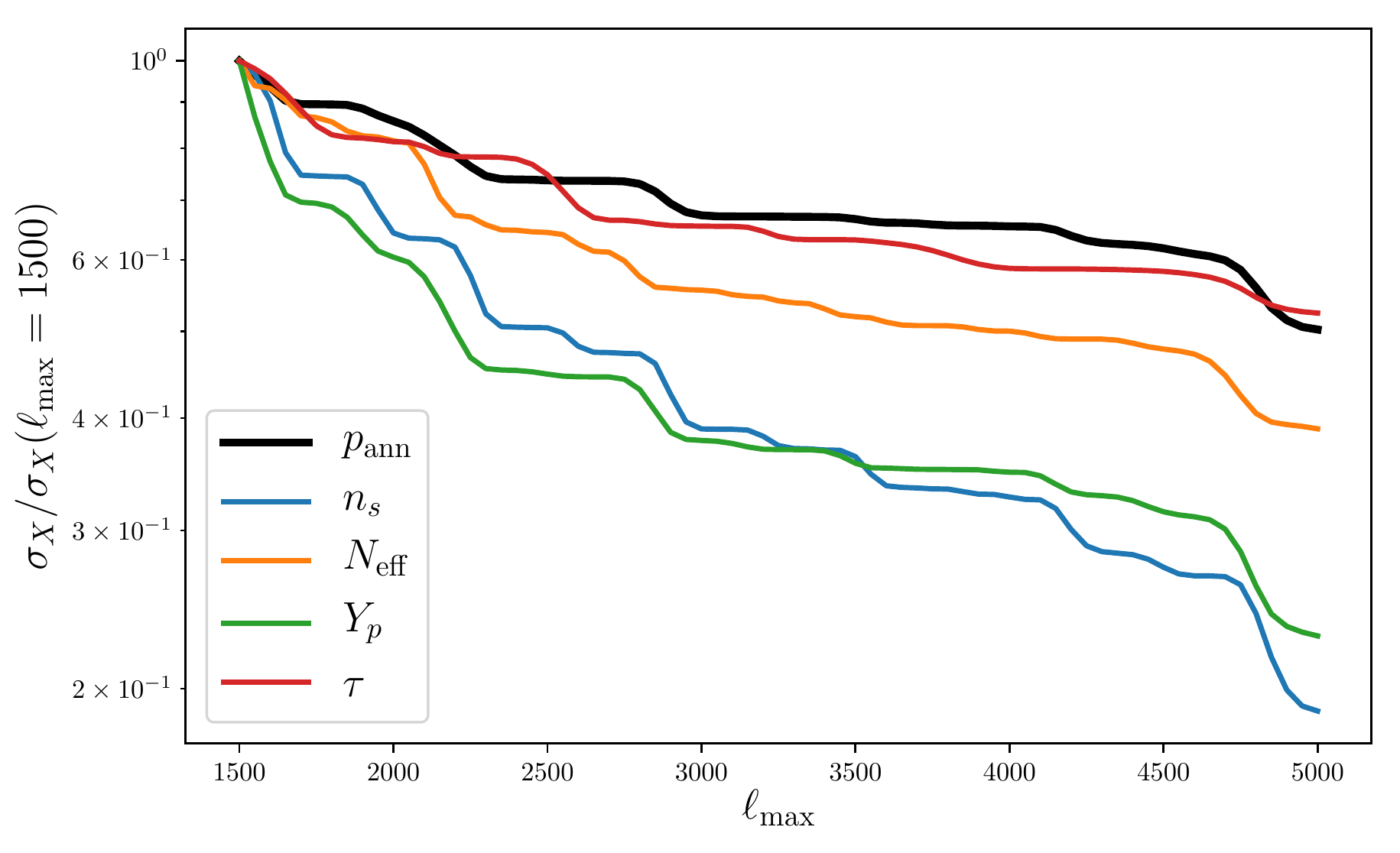}
\caption{ Improvement on forecasted constraints for a cosmic-variance-limited temperature-only CMB experiment up to a given $\ell_\mathrm{max}$ relative to that with $\ell_\mathrm{max}=1500$.}
\label{fig:fisher_normalized}
\end{centering}
\end{figure}

While the temperature and polarization spectra do indeed change at high $\ell$, forecasts for $\pann$ in Figures~\ref{fig:fisher_lmax} and~\ref{fig:fisher_normalized} show little similarity to $\Neff$ or $Y_p$.  Instead, the improvement (or lack thereof) is more similar to that of the reionization optical depth, $\tau$, than it is to other cosmological parameters.  

A priori, it is possible that the weak improvement of constraints on $\pann$ is due to a near-perfect degeneracy with another cosmological parameter. The phenomenological description in terms of a change to the width of recombination would suggest that a parameter like $\Neff$, $Y_p$, or even $n_s$\footnote{The spectral tilt $n_s$ does not lead to an exponential damping on small scales, but the effect of a tilt on the power spectrum is similar over a finite range in angular scale.} could exhibit such a degeneracy.  However, if  that were the case, we would expect both parameters to show little improvement and to have similar shapes in Figures~\ref{fig:fisher_lmax} and~\ref{fig:fisher_normalized}. The only parameter whose forecasts follow $\pann$ is $\tau$ which therefore suggests a very different origin for the $\pann$ constraint.  The lack of degeneracies with high-$\ell$ parameters is confirmed by a more complete exploration of the forecasts in Appendix~\ref{app:degen}.

\section{CMB Phenomenology of Dark Matter Annihilation}\label{sec:pheno}

\begin{figure}[t]
\begin{centering}
\includegraphics[width=1.005\columnwidth]{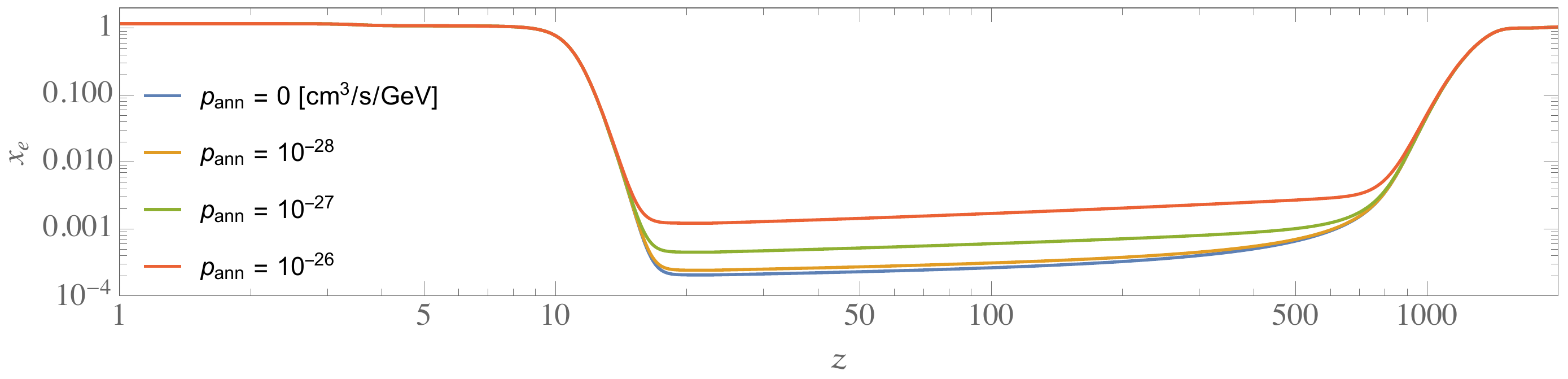}
\caption{Ionization histories for particular values of $p_\mathrm{ann}$.  The period between recombination and reionization exhibits the largest fractional change due to dark matter annihilation and is described by a non-zero value for $x_e$ which changes only weakly with time.}\label{fig:xe}
\end{centering}
\end{figure}

A lot of the intuition regarding the effect of dark matter annihilation on the CMB can be recovered from understanding the change to $x_e$, the ionization fraction.  The hydrogen ionization rate can be written as 
\beq
a^{-1}\dot{x}_e =\left(\beta_H e^{-\epsilon/T} (1-x_e) - \alpha_H x_e^2 n_e\right) C_H+ (I_\chi/n_e) \ , 
\label{eq:xebackground}
\eeq
where $\alpha_H$, $\beta_H$, and $C_H$ are defined in~\cite{Seager:1999bc} and describe conventional recombination and ionization.  The additional source of ionization from dark matter annihilation comes from $I_\chi \propto (1-x_e)$. Since $x_e \leq 1$ (by neutrality, assuming the helium is unionized) we see in Figure~\ref{fig:xe} that dark matter annihilation has its \emph{largest impact} on $x_e$ between recombination and reionization, when $x_e \ll 1$.

Since the dominant effect occurs when $x_e \ll 1$, we can simplify Equation~(\ref{eq:xebackground}) using $C_H = 1$, and in this limit the hydrogen ionization rate due to dark matter annihilation is
\beq
I_\chi \approx \frac{\pann \rho_\chi^2}{3\epsilon_H} \, ,
\eeq
where $\epsilon_H \simeq 13.6 \, \mathrm{eV}$ is the ionization energy of hydrogen.  Furthermore, the temperature after recombination is sufficiently small to neglect the contribution from $\beta_H$.  For annihilation rates of interest, the dark matter annihilation and recombination rates are much faster than the expansion rate, and one can find an approximate solution when these effects cancel~\cite{Padmanabhan:2005es,Dvorkin:2013cga}, giving
\beq
x_{e, {\rm floor}} \approx \frac{\rho_\chi}{\rho_b} \sqrt{\frac{1}{3(1-Y_p)^2} \frac{m_H^2}{\epsilon_H \alpha_H} \pann } \ .
\eeq 
For $z< 1000$, $\alpha_H \propto z^{-2/3}$ and so $x_{e, {\rm floor}} \propto z^{1/3}$.  Note that, even  though the annihilation rate is proportional to $\rho_\chi^2$ and therefore decreases like $a^{-6}$, the recombination rate is proportional to $n_e^2$ and therefore decreases at a similar rate. As a result, the effective ionization level only shows a weak dependence on redshift.

When $x_e \ll 1$, the visibility function can be well approximated by
\beq
g(\eta) = a n_e x_e \sigma_T \ .
\eeq
Decomposing $\Delta_{T,P}(\eta_0)\equiv \sum_\ell P_\ell(\mu) \delta\Theta^{T,P}_\ell$, the contribution from late times is given as 
\beq\label{eq:dT}
\delta \Theta^{T,P}_\ell \approx \int^{\eta_0}_{\eta_\star} \mathop{d\eta}  a n_e x_e \sigma_T s_{T,P}(\eta) j_\ell \!\left(k(\eta_0 - \eta)\right) \, .
\eeq
The universe was matter dominated between recombination and reionization such that $a \propto \eta^{2}$, $n_e \propto \eta^{-6}$, and $x_{e, {\rm floor}} \propto \eta^{-2/3}$.

\begin{figure}[t]
\begin{centering}
\includegraphics[width=0.49\columnwidth]{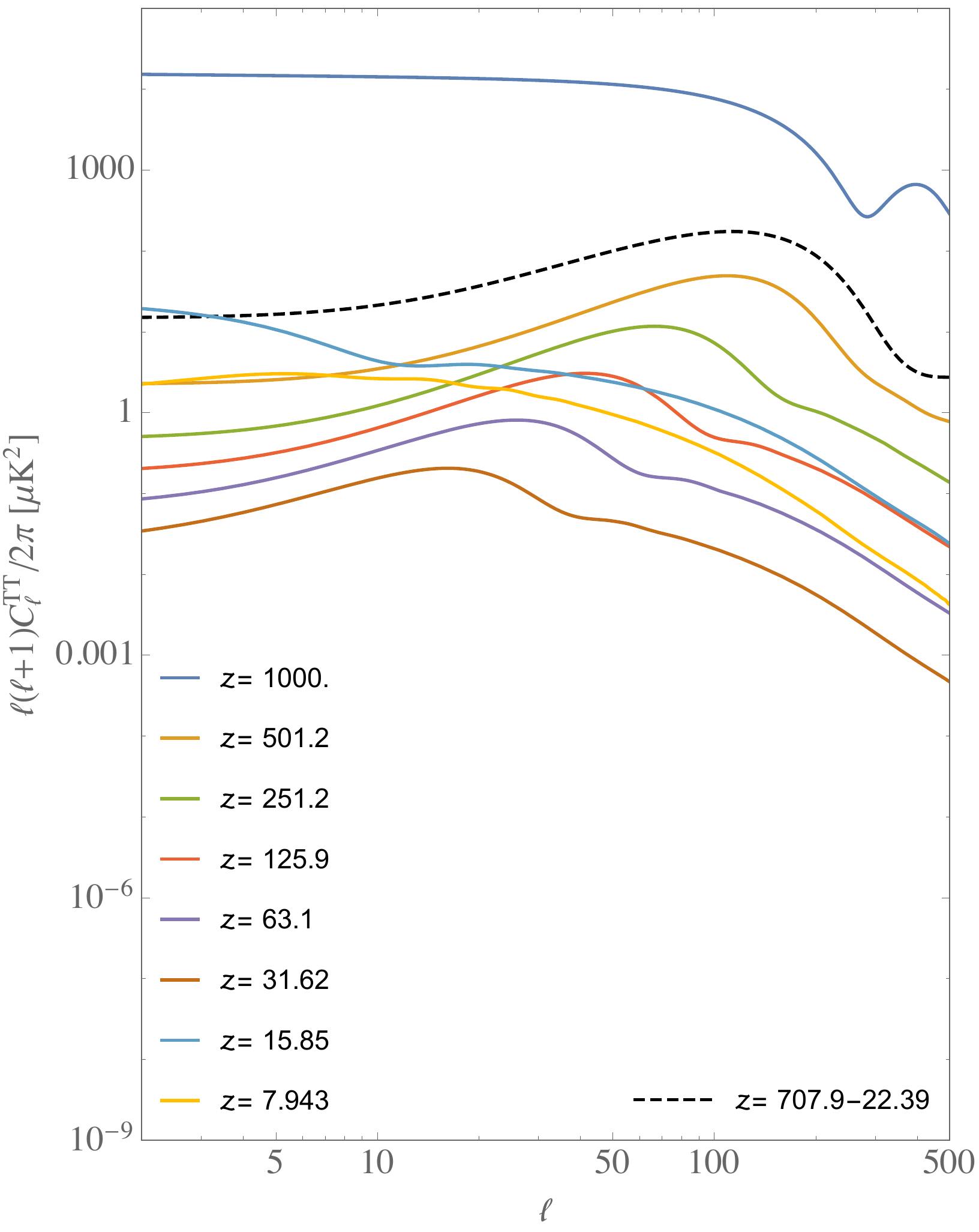}
\includegraphics[width=0.49\columnwidth]{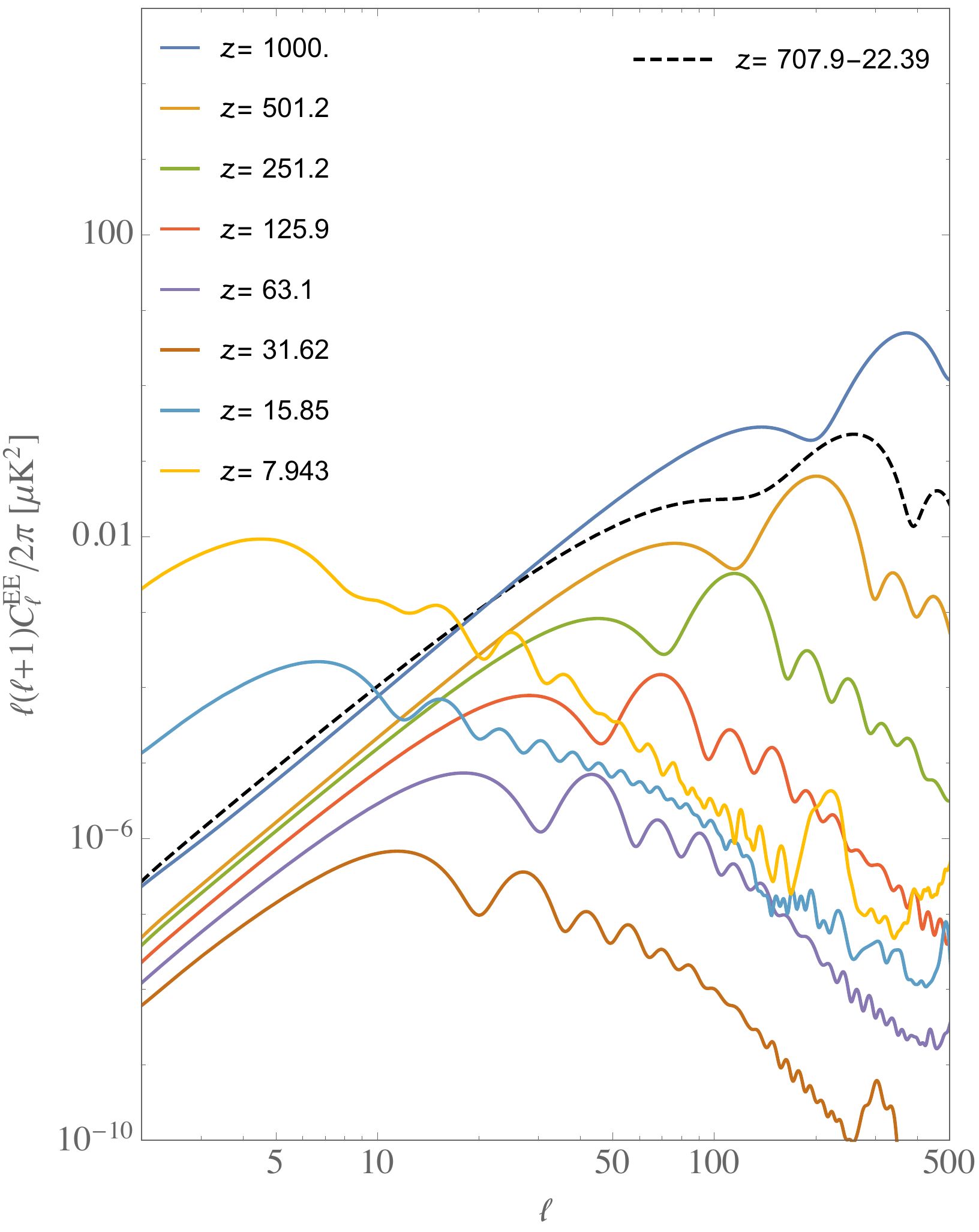}
\caption{Contributions to the temperature (left) and $E$-mode polarization (right) power spectra from various redshifts in a model with $p_\mathrm{ann}=10^{-26}$ cm$^3$/s/GeV.  Notice that the contributions from intermediate redshifts exhibit a peak between $\ell = 10$ and $\ell = 100$.  For temperature, the large scale power from recombination makes the large scale contributions from dark matter annihilation essentially unobservable.  However, for polarization, the sum of the power coming from redshifts between recombination and reionization results in an observable change to the spectrum which lies between the reionization bump and the first acoustic peak from recombination.}\label{fig:low_ell_z}
\end{centering}
\end{figure}

Let us first consider how the additional scattering between recombination and reionization (i.e.~increase in the visibility function) impacts the temperature anisotropy shown in the top panel of Figure~\ref{fig:EffectOfDMAonTTEEXe}. On large angular scales, $s_T \sim \Delta_{T,0}+\psi$ is approximately scale invariant and time-independent.  As a result, there is essentially no important time-dependence associated to the source function, and the contributions from late times in Equation~\eqref{eq:dT} are suppressed by $\eta^{-11/3}$.  This strong suppression at later times is largely a reflection of the dilution of the electron density from cosmological expansion and hence a corresponding reduction to the scattering cross-section.  In the left panel of Figure~\ref{fig:low_ell_z} we show the contributions to the temperature power spectrum from various redshifts.  One can see that the temperature power generated by late time scattering is subdominant compared to the contribution from recombination.  The most significant impact on the temperature power spectrum from the change to the visibility function is to suppress the high-$\ell$ power, 
much like the optical depth $\tau$. 

\begin{figure}
\begin{centering}
\includegraphics[width=0.85\columnwidth]{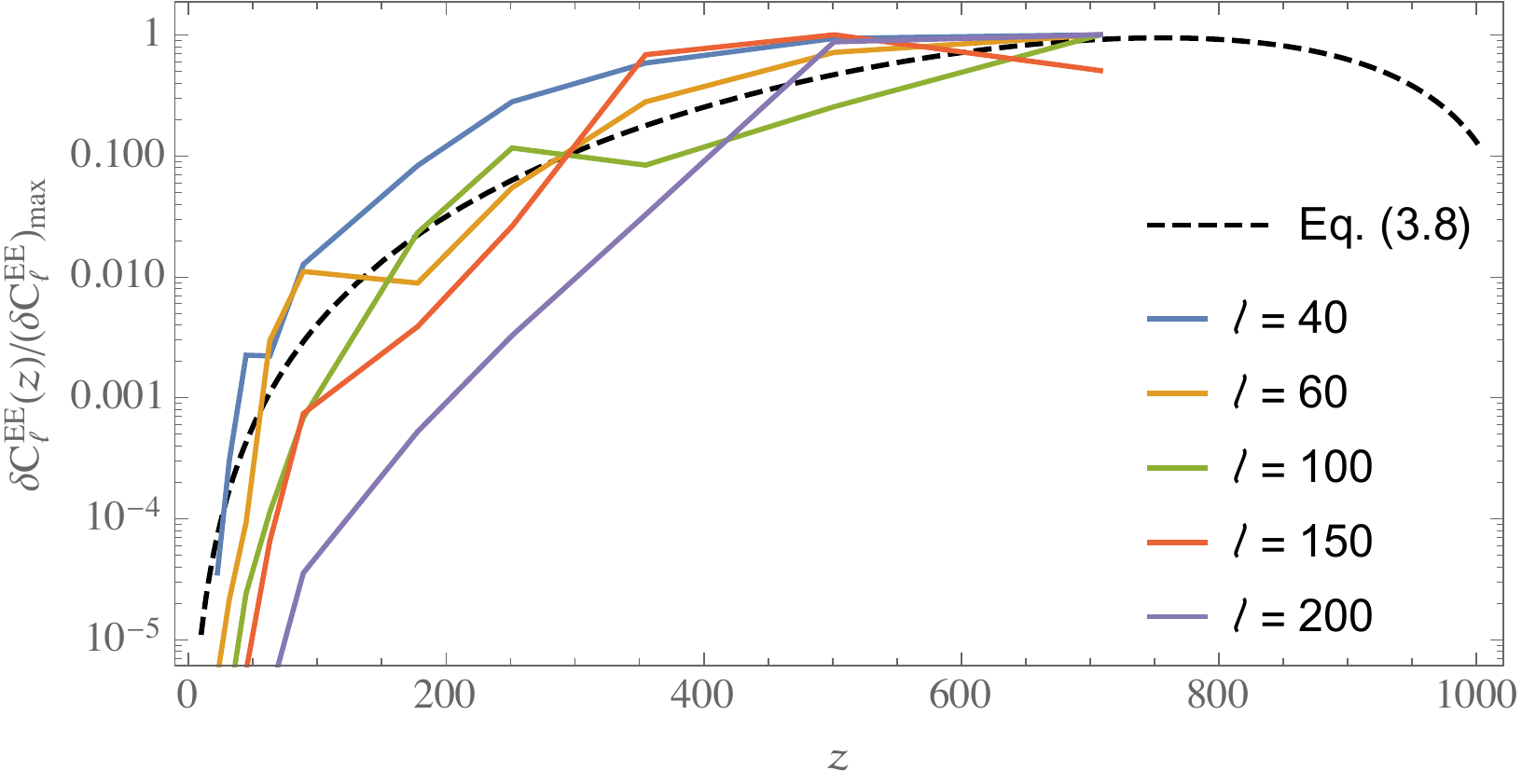}
\caption{The $E$-mode power contributed to specific angular scales by a range of redshifts in a cosmological model with $p_\mathrm{ann} = 10^{-26}$ cm$^3$/s/GeV, compared with that predicted by the analytical model described in Eq.~\eqref{eq:pol_contribs}.  }\label{fig:EE_z}
\end{centering}
\end{figure}

Now let us consider the case of the $E$-mode polarization shown in the bottom panel of Figure~\ref{fig:EffectOfDMAonTTEEXe}.  While the visibility function is essentially the same for polarization as for temperature, the source function is quite different.  At the time of recombination, the tight coupling approximation gives $\Pi \propto k\sin(k r_s) /\dot \tau(\eta_\star)$.  The amplitude of $E$-mode polarization from last scattering {\it at recombination} is therefore suppressed by $\dot{\tau}(\eta_\star) \gg 1$ and vanishes as $k^2$ in the limit $k\to 0$.  {\it After recombination}, the local temperature quadrupole grows from free-streaming giving
\beq
\Pi \approx \Delta_{T,2} = \Delta_{T,0}(\eta_\star) j_2 \! \left(k(\eta-\eta_\star)\right) \, .
\eeq
Using $k( \eta-\eta_\star) < 1$ we can approximate
\beq
\delta \Theta^{P}_\ell \approx \int^{\eta_0}_{\eta_\star} \frac{\mathop{d\eta}}{\eta} C k^2 (\eta-\eta_\star)^2 (\eta/\eta_r)^{-11/3} \Delta_{T,0}(\eta_\star) j_\ell \! \left(k(\eta_0 - \eta)\right) \, ,
\eeq
where $C = \frac{1}{15} a n_e x_e|_{\eta_r}$ and $\eta_r$ is some reference time between recombination and reionization.  In the Limber approximation, we anticipate that $k \approx \ell / (\eta_0 - \eta)$ and therefore the scaling with time is given by
\beq
\delta \Theta^{P}_\ell \propto \ell^2 \frac{(\eta-\eta_\star)^2}{(\eta_0 -\eta)^2} \eta^{-11/3} \Delta_{T,0} \, .
\label{eq:pol_contribs}
\eeq
Although this features the same $\eta^{-11/3}$ suppression we saw for the temperature fluctuations, the late time contribution to the $E$-mode polarization includes additional enhancements relative to the $E$ mode induced at recombination.  First, free-steaming means the observed polarization is sourced by a much larger temperature quadrupole (enhanced by $\dot \tau(\eta_\star)$).  Second, the low redshift scattering surface is closer to us and shifts the spectrum to lower $\ell$.  The right panel of Figure~\ref{fig:low_ell_z} shows how scattering which occurs between recombination and reionization leads to increased $E$-mode power on large scales, which is most visible  between the reionization bump and the first acoustic peak at recombination.  These competing effects explain\footnote{Some aspects of this model were also described in~\cite{Padmanabhan:2005es}.  Our description of the impact on the spectra is quantitative and matches the non-trivial redshift dependence of the signal.  We emphasize that this behavior is not consistent with simply changing the width of recombination.} the complicated redshift dependence seen in Figure~\ref{fig:EE_z}.

With this understanding in hand, we can now interpret the peculiar feature in the $E$-mode spectrum in the range $20 \lesssim \ell \lesssim 100$ in Figure~\ref{fig:EffectOfDMAonTTEEXe}.  Our analytic model explains why a change to $x_e$ in the redshift range $100 \lesssim z \lesssim 700$ can produce a larger $E$-mode power at low $\ell$ ($k \to 0$) than the $E$-mode power from recombination.  However, as shown in Figure~\ref{fig:low_ell_z} there is also the contribution from reionization ($z \lesssim 8$) which dominates over both of these contributions for $\ell \lesssim 20$.  As a result, the contribution from dark matter annihilation is most important in a narrow range of $\ell$, large enough to exceed the signal from reionization but small enough that the $E$ mode induced at recombination is sufficiently suppressed by the $k \to 0$ scaling.  Since this region reflects the large impact of $\pann$ on $x_e$ at intermediate redshifts, it explains why forecasts show that the CMB constraint on $\pann$ is essentially saturated by a cosmic-variance-limited measurement of $E$ modes on these scales.

\section{Impact on Large Scale Structure}\label{sec:lss}

\begin{figure}
\includegraphics[width=6.2in]{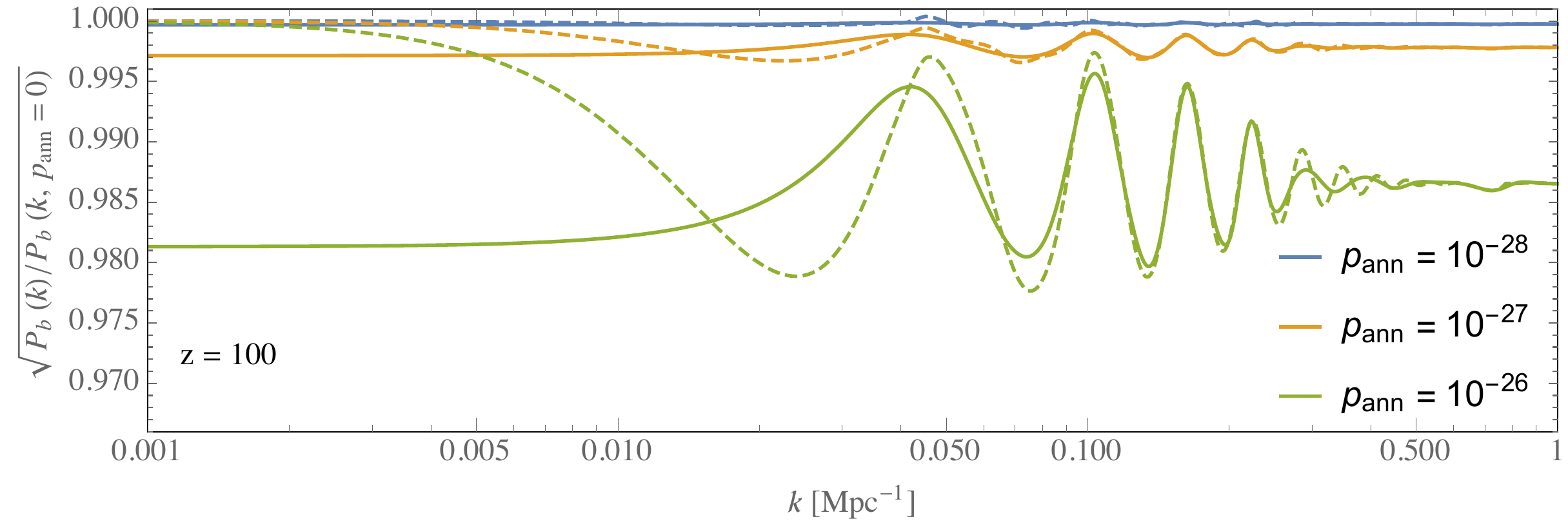}
\includegraphics[width=6.2in]{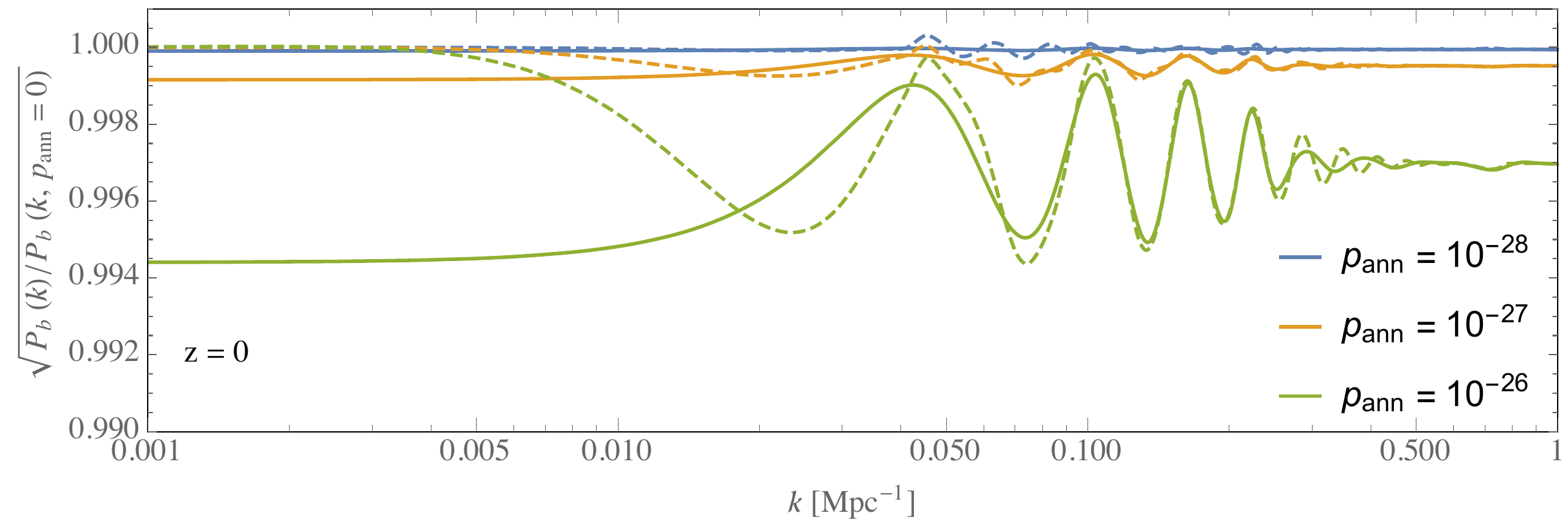}
\caption{Effect of dark matter annihilation $p_{\rm ann}$ on baryon fluctuations estimated using Eqs.~\eqref{eq:fluid1}-\eqref{eq:fluid5} at $z=100$ (top) and $z=0$ (bottom). The effect is largest at high redshift when photon drag is important. When the photon drag is neglected, the effect of dark matter annihilation is almost entirely described by a change to the initial conditions.  The dashed curves show the results from CAMB, finding good agreement on small scales confirming we identified all relevant effects in this regime. On large scales, our approximated fluid equations break down and we overestimate power as explained in the text.}
\label{fig:EffectOnPbar}
\end{figure}

Dark matter annihilation alters the post-recombination universe by heating baryons and increasing the ionization fraction.  While we have focused on the resulting effect on CMB anisotropies, the ionization fraction is also relevant for the evolution of structure in the later universe directly through its effect on baryons, which in turn couple to dark matter through the gravitational potential. In this section we will explore the resulting signatures of dark matter annihilation on large scale structure.

Dark matter annihilation affects both pre- and post-recombination physics. The dominant pre-recombination effect is a change to the size of the sound horizon which has an associated effect on the CMB (although it does not drive the constraints). Post-recombination, the increased ionization fraction alters the photon drag of the baryons and could, in principle, lead to observable changes in the growth of structure.

The dynamics of baryons are most relevant on small scales, well inside the horizon. We can therefore approximate their linear evolution in Fourier
space assuming the physical wavelength is much smaller than the horizon size. The evolution is described by the usual fluid equations in an expanding universe, with an additional advection term, i.e., 
\bea
&&\delta'_{c} +
\theta_c = 0 \, , \label{eq:fluid1}\\
&&\theta'_c + 2 H
\theta_c - \frac{k^2}{a} \phi = 0 \, ,~~~~\\
&&\delta'_{b}  + \theta_b = 0 \, ,\\
&&\theta'_b - 2 H \theta_b - \frac{k^2}{a^2} \phi - \frac{c_s^2}{a^2}
  k^2\delta_b -  \frac{4}{3} \frac{\rho_{\gamma}}{\rho_b}n_H x_e \sigma_T \theta_b =0 \, , \label{eq:baryon-momentum}\\
&&\frac{k^2}{a^2} \phi = -\frac{1}{2} \left(\rho_{b} \delta_b +
  \rho_{c} \delta_c\right) \, , \label{eq:fluid5}
\eea
where the subscripts $b$ and $c$ refer to baryons and cold dark matter,
respectively, and $\theta$ is the velocity divergence with respect to
proper space (following notation in \cite{Ali-Haimoud:2013hpa}). Derivatives with respect to proper
time are denoted by $' \equiv \partial_t$, while $\phi$ is the Newtonian gravitational potential.  Note that in the equations above we consider the unperturbed values of the speed of sound, the ionization fraction and the hydrogen number density. 
In Eq.~(\ref{eq:baryon-momentum}) $c_s$ is the average baryon isothermal
sound speed, which is related to the (mean) gas temperature $
c_s^2 \equiv \frac{T_{\rm gas}}{\mu ~m_{\rm H}}$. Although the gas temperature is affected by dark matter annihilation, we will show that this effect does not result in (observable) changes in the baryons and cold dark matter evolution. The mean molecular weight is also a function of the ionization fraction and thus could change the speed of sound in the presence of dark matter annihilation. However, this is a second order effect and 
for an essentially neutral plasma, the mean molecular weight is very nearly constant and with $Y_p \approx 0.25$, we have $\mu \approx 1.22$.  We have assumed fluctuations in the photon field velocity to be small. We also exclude {\it fluctuations} in the electron density and the baryon temperature, which would lead to small (higher order) corrections that will not concern us here.

\begin{figure}
\includegraphics[width=6.2in]{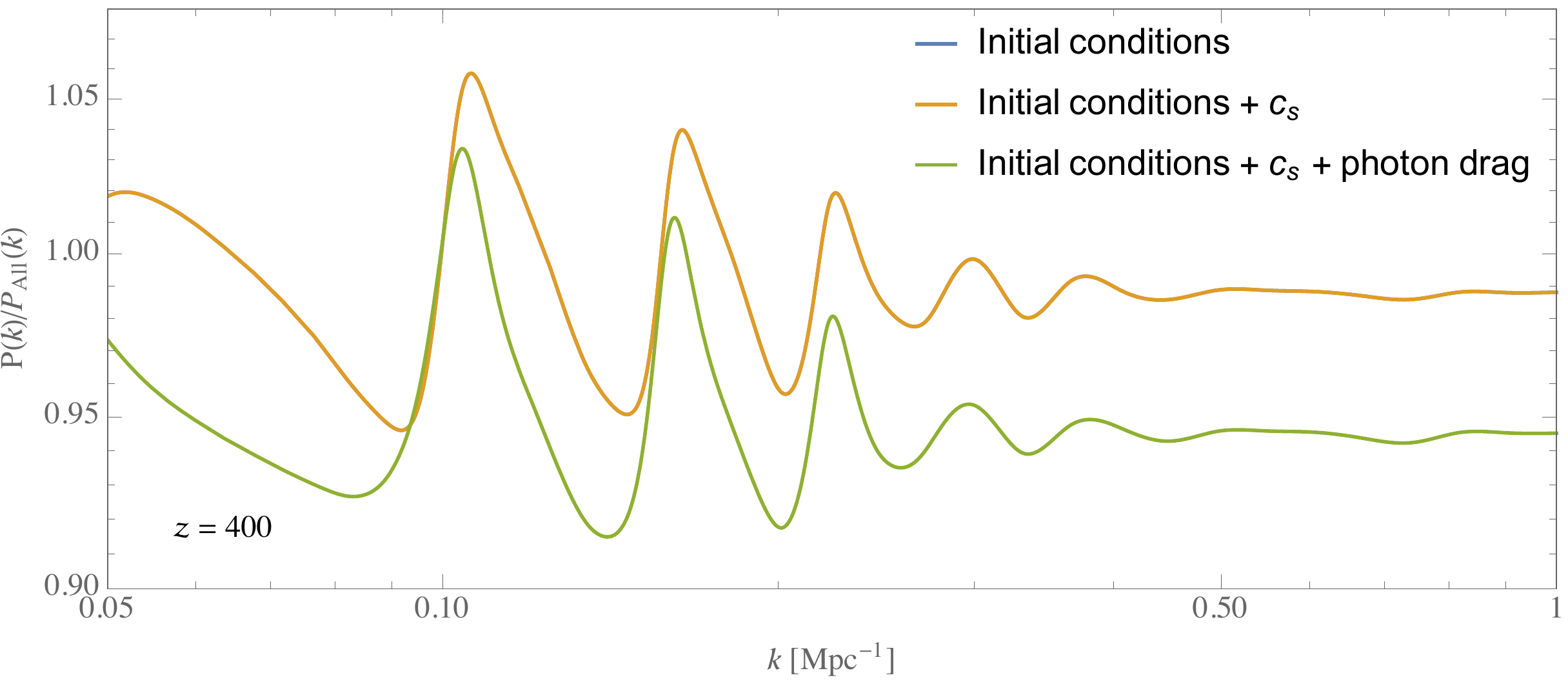}
\caption{The relative contribution as a result of dark matter annihilation from the 3 possible contributions in Eqs.~\eqref{eq:fluid1}-\eqref{eq:fluid5}. The effect is dominated by the initial offset at recombination and the photon drag term. Late-time changes to the sound speed do not cause observable effects (the blue line is almost entirely hidden by the orange line).}
\label{fig:EffectOnPbarPerTerm}
\end{figure}

Dark matter annihilation changes the size of the sound horizon around recombination, which shifts the peaks of the baryon acoustic oscillations in the matter power spectrum and the CMB.  This effect could be observed independently from the post-recombination effects we will discuss next.  Although the sound speed also plays a role in the dynamics at late times through the pressure term, this will not have large effects on the matter power spectrum. On the other hand, the photon drag term (the last term in Eq.~\eqref{eq:baryon-momentum}) cannot be ignored. We will see that this term introduces a suppression of power.

We numerically solve the differential equations above using initial conditions set by CAMB~\cite{Lewis:1999bs} at $z = 900$. We show the effect on (linear) baryon fluctuations in Figure~\ref{fig:EffectOnPbar}. {\it The photon drag term leads to a relatively large effect on the fluctuations at early times.} As the drag term becomes small compared to the Hubble rate, the amplitude of the effect is reduced by gravitational coupling with the cold dark matter (which is not affected at early times, apart from the shift in the baryon acoustic oscillations due to change to the sound horizon). Today, the effect is sub-percent. For $\pann =3.4\times 10^{-28} \, {\rm cm}^3/{\rm s}/{\rm GeV}$, the maximum currently allowed by the Planck data~\cite{Ade:2015xua}, the effect is of the order of a tenth of a percent on the amplitude of fluctuations. This has to be compared to, for example, the effect on the large scale $E$-mode polarization spectrum, which for the same value of $\pann$ is modified by almost ten percent. 

In Figure~\ref{fig:EffectOnPbarPerTerm} we show the effect of the initial conditions (at recombination), the sound speed, and the photon drag on the matter power spectrum at $z = 400$ for $\pann=10^{-26}\, {\rm cm}^3/{\rm s}/{\rm GeV}$.  The initial conditions determine the offset of the acoustic oscillations.  The sound speed has no visible effect, not even on the baryons. The photon drag, on the other hand, is crucial to explain the overall suppression of power. 

To better understand the influence of the photon drag let us take a closer look at Eq.~\eqref{eq:baryon-momentum}. The photon drag is typically ignored at late times, because it is subdominant to the Hubble expansion, i.e. 
\bea
2H \gg   \frac{4}{3} \frac{{\rho}_{\gamma}}{{\rho}_b}n_H x_e \sigma_T \, . 
\eea
In a matter-dominated universe $H \propto z^{2}$ while the right-hand side scales as $z^4 x_e(z)$. Since ${x}_e(z) \propto z^n$ with $n > 0$ before reionization, the Hubble expansion quickly overtakes photon drag, even though it dominates at early times. Numerically we find the photon drag becomes sub-dominant at $z \simeq 900$; it is also the moment at which the relative effect, compared to no annihilation, is largest. For the largest values of $\pann$ we have considered in this paper we find that the change in the effective drag term (i.e. the term multiplying $\theta_b$) is of order $8\%$ at $z = 900$ which would result in a modulation of the power spectrum of baryons of order $15\%$ 
if the baryon power spectrum would be determined entirely by this term. Taking into account the other terms in Eq.~\eqref{eq:baryon-momentum} and the fact that the modulation is reduced due to gravitational interaction with cold dark matter, the total effect is smaller. 

\begin{figure}
\includegraphics[width=6.2in]{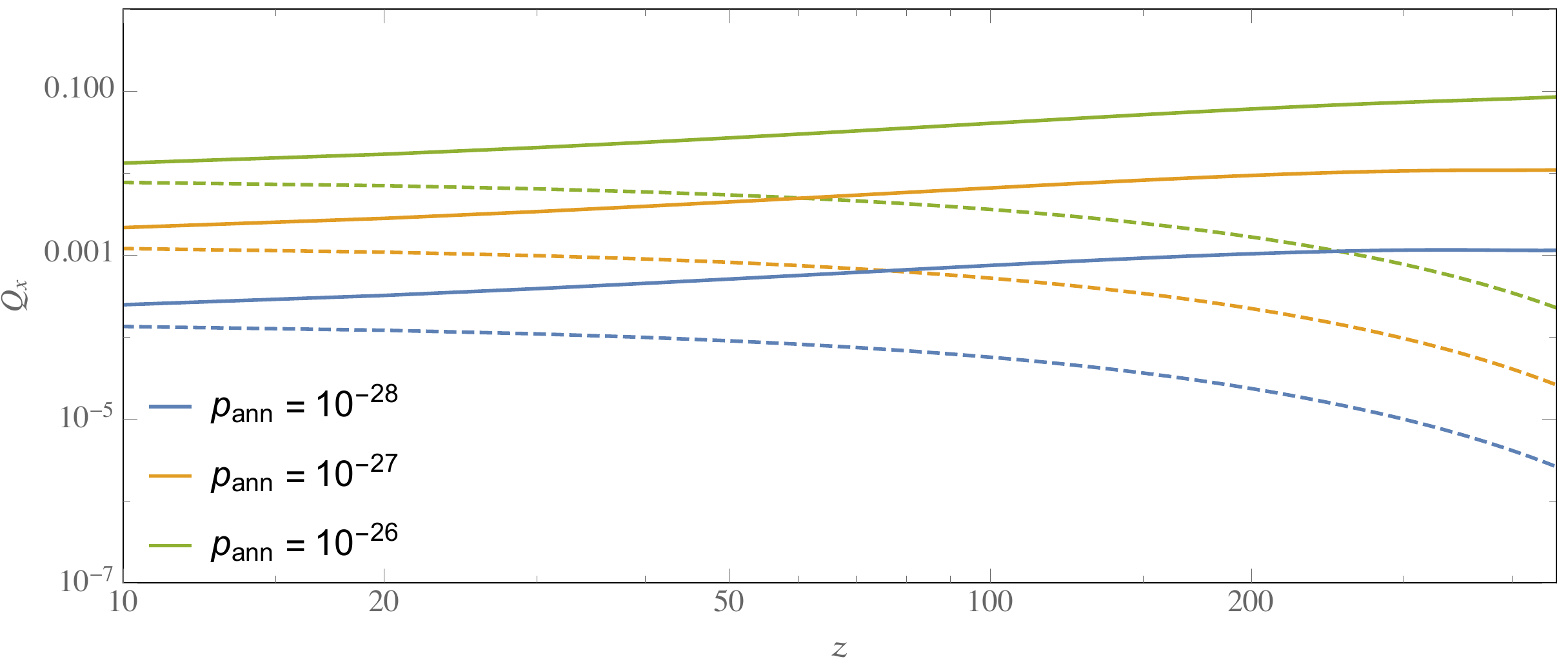}
\caption{Effect of dark matter annihilation $p_{\rm ann}$ on baryon (solid) and dark matter (dashed) fluctuations as function of redshift measured by Eq.~\eqref{eq:relchange}. Although effects on baryons are relatively large at high redshift, effects are suppressed at late times due to gravitational coupling to dark matter.}
\label{fig:Meanzeffect}
\end{figure}

In Figure~\ref{fig:Meanzeffect} we aim to qualify the effect on the baryons compared to the effect on the cold dark matter as a function of redshift. We plot the dimensionless integrated relative change with respect to the case with no dark matter annihilation for both baryons and cold dark matter:
\bea
Q_x(z) \equiv  \int \frac{\mathop{dk}}{k} \left| \left(\frac{P_x(k)}{P_x(k,\pann= 0)}\right)^{1/2}-1\right| \, , \label{eq:relchange} 
\eea
with $x$ representing either cold dark matter or baryons. We run the integral from $0.05 \leq k \leq 1$ Mpc$^{-1}$. As expected, Figure~\ref{fig:Meanzeffect} shows  that the effect on the baryons is largest at high redshift while dark matter annihilation hardly changes the fluctuations in the cold dark matter. The figure also illustrates clearly that the effect on the baryons is suppressed at late times as gravitational coupling with the dark matter washes out the effect.

Although the effect is small, the analysis shows that the impact on the power spectra appears on all scales $k\geq 0.05$ Mpc$^{-1}$. With tomography of the power spectrum it might be possible to constrain this small suppression. Since the baryons are much more affected than the cold dark matter at high redshift, an experiment which traces baryons rather than dark matter would be more sensitive to the effect of dark matter annihilation. Mapping the 21cm signature coming from neutral hydrogen provides one such possibility. Although obviously not an unbiased tracer of the baryon field, it would present the cleanest signature at high redshifts. We will not make an attempt to estimate the detectability here, since no experiments are currently planned to measure the 21cm signal at such high redshifts (see e.g. Refs.~\cite{Furlanetto:2006wp, Chuzhoy:2007fg}).  Dark matter annihilation also heats the gas, which affects the brightness temperature of the 21cm signal. This effect would likely be much easier to observe, and several studies have shown that dark matter annihilation could be constrained using the global 21cm signal~\cite{Cumberbatch:2008rh,Valdes:2012zv}.

At the same time, on small scales, as computed here, the effect will be hard to distinguish from overall modulations of the power and we expect it would require a very careful measurement of the primordial amplitude of fluctuations to estimate effects from dark matter annihilation comparable to the effects on small scales from massive neutrinos. 

Note also that we have limited ourselves to the linear matter power spectrum only. At late times, non-linear effects become more important and it has been shown that dark matter annihilation would alter the matter power spectrum through enhanced annihilation in collapsed objects \cite{Giesen:2012rp}. These effects might be larger, but also require more assumptions than our simple treatment here.

Thus far, we have only discussed the effects of dark matter annihilation on scales which are small compared to the horizon. Figure~\ref{fig:EffectOnPbar} shows how our approximation breaks down compared to the full numerical results from CAMB on large scales. The difference is a consequence of underestimating the total power on large scales. Solving the fluid equations we dropped terms coming from the time derivative of the large scale potential, i.e. $\phi'$ in Eq.~\eqref{eq:fluid1} and velocity in the equation for the potential Eq.~\eqref{eq:fluid5}. As a result our power spectrum on large scales will be suppressed compared to results from the complete set of equations. Since we measure the relative effect, an underestimate of the power on large scales increases the relative contribution. However, neither of the neglected terms is affected by $\pann$ and as such we do not expect to be missing physical effects that could change the power spectrum on large scales {\it as a result of dark matter annihilation}. 
Conclusions drawn based on our small scale results should hold.

\section{Summary}\label{sec:sum}

Dark matter annihilation is a compelling signature of WIMP dark matter.  Current CMB observations provide a powerful constraint, particularly at low mass, that has proven to be a valuable window into the nature of dark matter.  In this paper, we explored the physical origin of the cosmological bounds on dark matter annihilation. The primary impact is on the ionization fraction and can only change substantially at times when the universe was neutral.  We demonstrated using a simple model that the largest impact on the CMB arises from the change to the ionization fraction at redshifts $z \sim 100-700$ with a predictable scaling.  The phenomenology is physically similar to the optical depth to reionization, $\tau$, and shows similar behavior in forecasts for future CMB experiments.  This explains why current constraints on $\pann$ have nearly saturated the ultimate cosmic variance limit.

A better understanding of the CMB constraint is important for interpreting existing bounds on $\pann$ as a constraint on specific models.  The redshift dependence of the signal is non-trivial and peaks at lower redshifts than the naive expectations.  While such redshift dependence is captured in Boltzmann codes, one would like to apply the bound without having to recalculate the spectra for every possible model.  Furthermore, a physical understanding of the origin of the constraints can help identify blind spots in the inferences drawn from CMB analyses and suggest new targets for dark matter searches.  We hope the phenomenological description provided in this paper will be a useful tool in the quest to understand the nature of the dark matter.

Given the theoretical importance of constraining $\pann$, it is natural
ask if other cosmological observables can circumvent the limitations of the conventional CMB analysis.  We explored one possibility, namely the impact of dark matter annihilation on large scale structure after recombination.  In addition, there are a variety of  effects on the CMB that could provide additional windows into the nature of dark matter that are either not captured by $\pann$ or appear as higher order corrections.  Spectral distortions are the most common such example \cite{Ali-Haimoud:2015pwa}, but other possibilities include the effect on the CMB from a dark matter annihilation-induced reionization history \cite{Giesen:2012rp}, higher order effects on the recombination history \cite{Dvorkin:2013cga}, and impact\footnote{Because of the recent EDGES measurement claiming a detection of the global 21cm signal~\cite{EDGES2018} with an absolute brightness temperature too high to be explained using minimal modeling of the spin temperature, several papers have appeared that derive rather tight constraints on dark matter annihilation (see e.g.~\cite{DAmico:2018sxd,Liu:2018uzy}).  The constraining power of the global signal depends sensitively on the (still uncertain) central value of this temperature measurement and it remains to be seen if it will ultimately exceed the sensitivity of the CMB.} on the global 21cm signal~\cite{Cumberbatch:2008rh,Valdes:2012zv}. 

More broadly, there are a wide range of astrophysical searches for the annihilation products of dark matter; see~\cite{Gaskins:2016cha} for a review.  While there have been several tantalizing hints from these searches, none have yet resulted in a convincing detection.  These different approaches are important complements to our understanding of dark matter, as cosmology places broad constraints on the total dark matter annihilation cross-section and these astrophysical searches are capable of determining the specific annihilation channels of dark matter.

Cosmology offers a variety of windows into the nature dark matter beyond the annihilation signal alone.  Laboratory-based constraints already hint that the traditional WIMP may not be the answer and new and more powerful windows may be needed to expose its identity.  This should motivate further work applying the growing sensitivity in cosmological observations to this fundamental problem.

\acknowledgments

We wish to thank Yacine Ali-Haimoud and Raphael Flauger for useful discussions.  D.\,G.~thanks the University of California, Berkeley for hospitality while this work was being completed.
P.D.M. acknowledges support from a Senior Kavli Fellowship at the University of Cambridge and support from the Netherlands organization for scientific research (NWO) VIDI grant (dossier 639.042.730).

\newpage

\clearpage
\appendix

\section{CMB Forecasts and Degeneracies}\label{app:degen}

\begin{table}[t!]
\begin{center}
 \begin{tabular}{lll } 
 \toprule
   Parameter   &   Fiducial Value      & Step Size     \\ [0.5ex] 
 \midrule
   $\Omega_c h^2$ 		&   0.1197 	            & 0.0030 	    \\ 
   $\Omega_b h^2$ 		&   0.0222 	            & $8.0\times10^{-4}$ 	    \\
   $\theta_s$     		&   0.010409 	        & $5.0\times10^{-5}$ 	    \\
   $\tau$         		&   0.060 	            & 0.020 	    \\
   $A_s$          		&   $2.196\times10^{-9} \quad $  & $0.1\times10^{-9}$ 	    \\
   $n_s$          		&   0.9655 	            & 0.010 	    \\
   $\sum m_\nu$ [eV]	&   0.060 	            & 0.020 	    \\
   $\Neff$          	&   3.046 	            & 0.080 	    \\
   $Y_p$          		&   0.2467 	            & 0.0048 	    \\
   $\pann \,[10^{-27} \mathrm{cm}^3/\mathrm{s}/\mathrm{GeV}]  \quad $	&	0.0	& 0.06	\\
  \bottomrule
\end{tabular}
\caption{Fiducial cosmological parameters and step sizes for numerical derivatives used in forecasts.  All forecasts use the 10-parameter $\Lambda$CDM+$\sum m_\nu$+$\Neff$+$Y_p$+$\pann$ model even when only a subset of parameters are shown.  There is no significant change in the behavior of forecasts for $\pann$ with a different choice of cosmology.  All numerical derivatives are computed using a centered difference, except for $\pann$ for which we use a forward difference, since negative values of $\pann$ are undefined.
}
\label{table:cosmo_fiducial}
\end{center}
\end{table}

In this appendix, we provide more information regarding the forecasts presented in the main text.  We forecast constraints by computing the Fisher matrix defined as \cite{Fisher1935,Knox:1995dq,Jungman:1995bz}
\beq
	F_{ij} = \sum_\ell \frac{2\ell + 1}{2} f_\mathrm{sky} \mathrm{Tr} \left(\mathbf{C}_\ell^{-1} \frac{\partial \mathbf{C}_\ell}{\partial \lambda_i} \mathbf{C}_\ell^{-1} \frac{\partial \mathbf{C}_\ell}{\partial \lambda_j} \right) \, ,
\eeq
with the covariance matrix $\mathbf{C}_\ell$ given by
\beq 
	\mathbf{C}_\ell = 
    \begin{bmatrix}
    C_\ell^{TT} & C_\ell^{TE} & 0 \\
    C_\ell^{TE} & C_\ell^{EE} &  0 \\
    0 & 0 &  C_\ell^{dd}
	\end{bmatrix}
    +\mathbf{N}_\ell \label{eq:covariance}
\eeq
and the noise covariance is taken to be diagonal $\mathbf{N}_\ell = \mathrm{diag}\left( N_\ell^{TT}, N_\ell^{EE}, N_\ell^{dd}\right)$.  For the cosmic-variance-limited forecasts, we take the temperature, polarization, and lensing reconstruction noise to vanish and use unlensed spectra.  For the forecasts presented in Figure~\ref{fig:forecasts}, the noise is assumed to be Gaussian with
\beq
	N_\ell^{TT} = \Delta_T^2 \exp \left(\ell(\ell+1)\frac{\theta_{\mathrm{FWHM}}^2}{8\log 2}\right) \, , 
\eeq
where $\Delta_T$ is the instrumental noise in $\mu$K-radians and $\theta_{\mathrm{FWHM}}$ is the beamsize in radians.  The polarization noise is given by an equivalent expression with $\Delta_P=\sqrt{2} \Delta_T$. The lensing reconstruction noise is calculated using a minimum variance quadratic estimator~\cite{Hu:2001kj}, including the improvement from iterative delensing with the $EB$ estimator~\cite{Smith:2010gu}.  We use spectra which are internally delensed in order to improve parameter constraints~\cite{Green:2016cjr}.

To calculate power spectra, we use the Boltzmann code CAMB~\cite{Lewis:1999bs}, with dark matter annihilation included through a modified version of the recombination code HyRec~\cite{AliHaimoud:2010dx}.  The fiducial values of the cosmological parameters and step sizes used for numerical derivatives are shown in Table~\ref{table:cosmo_fiducial}.  

\begin{figure}
\begin{centering}
\includegraphics[width=0.7\columnwidth]{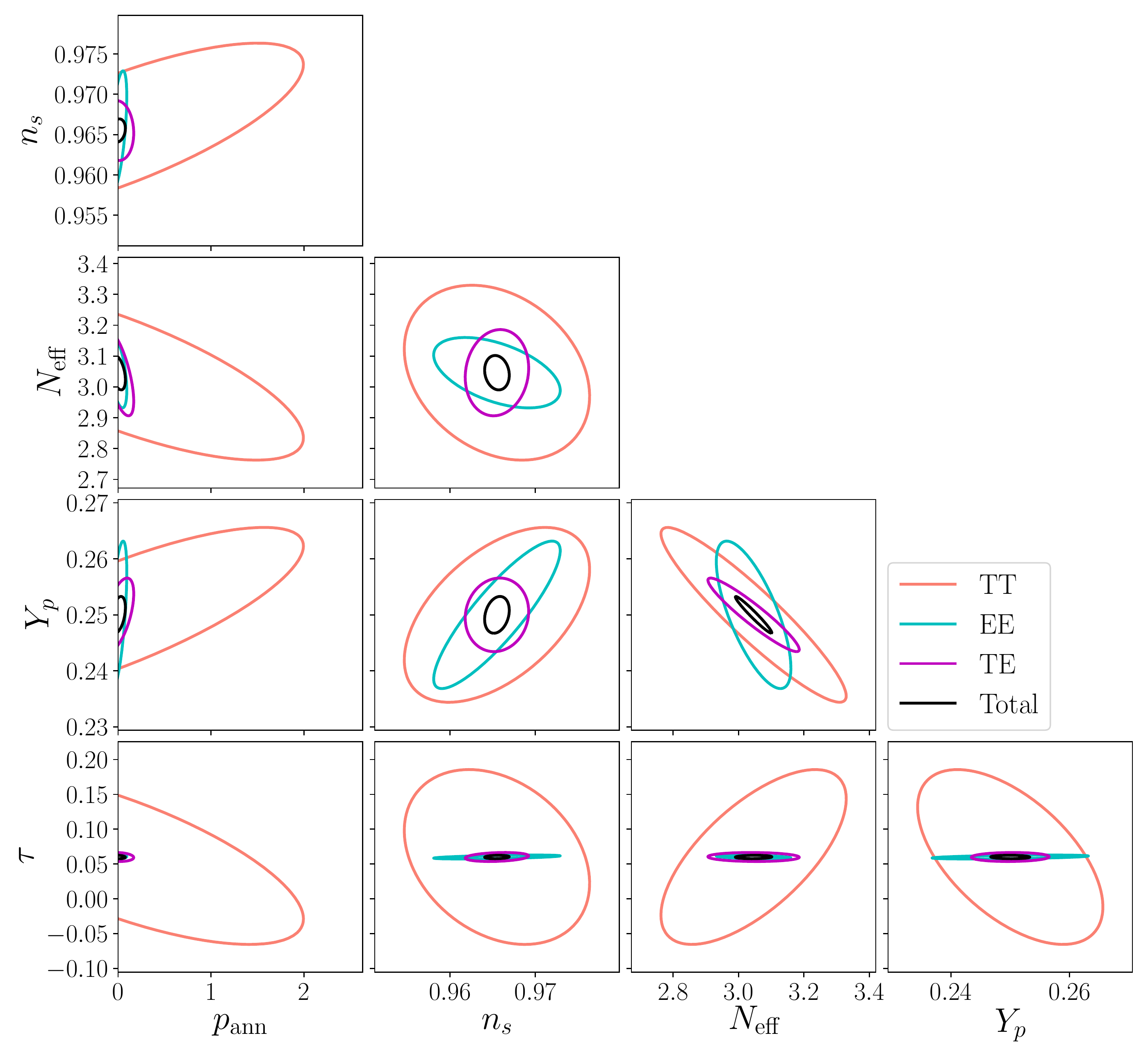}
\caption{ Forecasts showing two-dimensional 1-$\sigma$ constraints for selected parameters for a cosmic-variance-limited CMB experiment with $\ell_\mathrm{max} = 5000$.  The constraints on $\pann$ are shown in units of $10^{-27} \mathrm{cm}^3/\mathrm{s}/\mathrm{GeV}$.  Notice there is only mild degeneracy present among the parameters when including only temperature data which nearly disappears for $p_\mathrm{ann}$ when including polarization data.}
\label{fig:fisher_ellipses_CV}
\end{centering}
\end{figure}

Figure~\ref{fig:fisher_ellipses_CV} shows the forecasted 1-$\sigma$ two-dimensional contours for the same set of parameters shown in the main text. Notice that constraints on $\pann$ are not limited by any significant degeneracy, even when using only temperature information from the CMB.  When polarization is included, the mild degeneracies that do exist are even further suppressed.  This highlights the point that the constraints on $\pann$ are dominated by polarization data at relatively large angular scales, which explains why CMB constraints on dark matter annihilation are mostly saturated with existing data.

\addcontentsline{toc}{section}{References}

\bibliographystyle{utphys}

\bibliography{dm_refs}

\end{document}